基金项目：吉林省自然科学基金项目(YDZJ202101ZYTS149)
项目名称：面向系统韧性提升的综合能源系统优化调度研究
起止时间：2021 年 1 月-2023 年 12 月　　　　　项目负责人：李扬
Project supported by Natural Science Foundation of Jilin Province (YDZJ202101ZYTS149).

# 多智能体深度强化学习驱动的跨园区能源交互优化调度

李扬[1]，马文捷[1]，卜凡金[2]，杨震[3]，王彬[4]，韩猛[2]

(1.东北电力大学，吉林省吉林市 132012；
2.国网淄博供电公司，山东省淄博市 255022；
3.国网北京市电力公司，北京市 100032；
4.国网济宁供电公司，山东省济宁市 272000)

**摘要**:为协调多园区综合能源系统各个园区之间的能量交互，多能源子系统之间的能源转换，实现综合能源系统整体优化调度，本文提出一种利用多智能体深度强化学习算法学习不同园区的负荷特征，并在此基础上进行决策的综合调度模型。该模型将多园区综合能源系统的调度问题转化为马尔科夫决策过程，并利用深度强化学习算法进行求解，避免了对多园区、多能源子系统之间复杂的能量耦合关系进行建模。仿真结果表明，所提方法可以很好的捕捉到不同园区的负荷特性，并利用其中的互补特性协调不同园区之间进行合理的能量交互，可以实现弃风率由 16.3%降低至 0，并可以使总运行成本降低 5445.6 元，具有良好的经济效益和环保效益。

**关键词**: 多智能体深度强化学习；综合能源系统；优化调度；可再生能源消纳；负荷特征学习;多园区能量交互

## Deep Reinforcement Learning-driven Cross-Community Energy Interaction Optimal Scheduling

LI Yang[1], MA Wenjie[1], BU Fanjin[2], YANG Zhen[3], WANG Bin[4], HAN Meng[2]

(1. Northeast Electric Power University, Jilin 132012, China
2. State Grid Zibo Power Supply Company, Zibo 255022, China
3. State Grid Beijing Electric Power Company, Beijing 100032, China
4 State Grid Jining Power Supply Company, Jining 272000, China)

**ABSTRACT:** In order to coordinate energy interactions among various communities and energy conversions among multi-energy subsystems within the multi-community integrated energy system under uncertain conditions, and achieve overall optimization and scheduling of the comprehensive energy system, this paper proposes a comprehensive scheduling model that utilizes a multi-agent deep reinforcement learning algorithm to learn load characteristics of different communities and make decisions based on this knowledge. In this model, the scheduling problem of the integrated energy system is transformed into a Markov decision process and solved using a data-driven deep reinforcement learning algorithm, which avoids the need for modeling complex energy coupling relationships between multi-communities and multi-energy subsystems. The simulation results show that the proposed method effectively captures the load characteristics of different communities and utilizes their complementary features to coordinate reasonable energy interactions among them. This leads to a reduction in wind curtailment rate from 16.3% to 0% and lowers the overall operating cost by 5445.6 Yuan, demonstrating significant economic and environmental benefits.

## 0 引言

随着化石能源的枯竭，构建可以充分利用可再生能源（Renewable Energy, RE）的综合能源系统（Integrated Energy System，IES）成为一种积极的选择[1]。关于 IES 相关的示范工程建设，欧洲的框架计划 FP6 致力于含分布式可再生能源发电的微电网研究[2]，并在丹麦的 Bornholm 岛上构建了一个较大规模的 IES，该 IES 中包含热电联产机组、分布式风力和光伏发电机、柴油发电机组以及沼气发电机，通过海底电缆与瑞典电网相连，可以孤立运行也可以并网运行[3]。我国在东部沿海构建了一些 IES 示范园区，有力推动了 IES 的推广应用[4]。IES 通过整合不同能源形式，打破了传统能源子系统的运行限制，以提高能源利用效率，因此其优化运行研究成为当前研究的热点[5-6]。

关于 IES 的优化运行，目前已有较多的相关研究。文献[7]构建风机、光伏发电的出力概率模型，然后针对 RE 的出力情况在一定的置信度下设置旋转备用，这样可以达到运行经济性与可靠性的折中。文献[8]采用不确定集合表征的方法，以区间形式描述风速数据，构建双层鲁棒模型。文献[9]利用梯形模糊数来表示风机、光伏发电的出力来构建 IES 的调度模型。上述研究多数是基于对 IES 内部耦合关系的精确建模和对 RE 出力的精确预测，在含多种能量耦合交互的 IES 中实现此类精确建模和预测是十分困难的。

为减少对能量耦合关系建模和预测信息的依赖，近年来有学者将基于数据驱动的深度强化学习（Deep Reinforcement Learning，DRL）方法引入到能量管理系统中。DRL 可以通过与外部环境的互动自动地感知外部环境的特征和变化并对应地调整策略，避免了对多重复杂不确定性的建模。文献[10]将深度神经网络与 Q 学习相结合，获得了一个对电网优化管理的方法—深度 Q 网络（Deep Q Network, DQN）。文献[11]中的 DQN 通过对历史电价的学习，调动蓄电池的充放电，实现了对微电网负荷的调整。文献[12]利用训练后的 DQN 处理微电网与外部电网的能量交换。文献[13]利用柔性行动器-判别器（Soft Actor-Critic，SAC）算法以实时电价和电动汽车的剩余电量作为探索环境，获得最佳的充电策略。利用基于 SAC 的恒温负荷智能分区管理可以提高 RE 利用率和用户的热舒适性[14-15]。然而，现有研究考虑的能源耦合形式比较少且多数为单园区系统，与当前实际运行的多园区、多能源耦合的 IES 运行形式不相符。

本文的主要创新点可以总结如下：

（1）构建了考虑多能耦合的多园区 IES 模型，对其中的关键设备进行了详细建模，并建立了描述多园区之间能量交互的模型。

（2）提出了以最小成本为目标的多园区 IES 调度模型，并将该调度模型转化为 DRL 框架下的马尔科夫决策过程（Markov Decision Processes, MDP），避免了对复杂的多能耦合关系进行建模的需要。

（3）利用多智能体深度强化学习算法学习并利用不同园区的负荷特征，以有效协调多园区之间的能量交互，实现 IES 的整体优化，从而能够更好地应对多园区、多能源耦合的 IES 的运行特性。

## 1 多园区综合能源系统建模

### 1.1 系统结构

多园区 IES 是指在同一 IES 中，不同的产业集群在不同的地理空间上聚集，出现了不同的产业园区[16]。

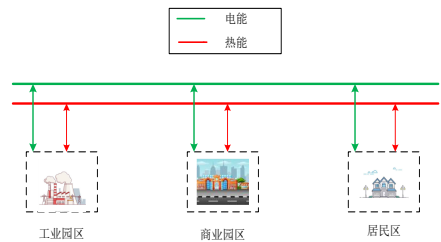

**图 1 多园区综合能源系统结构图**

**Fig. 1 Structure of multi-community IES**

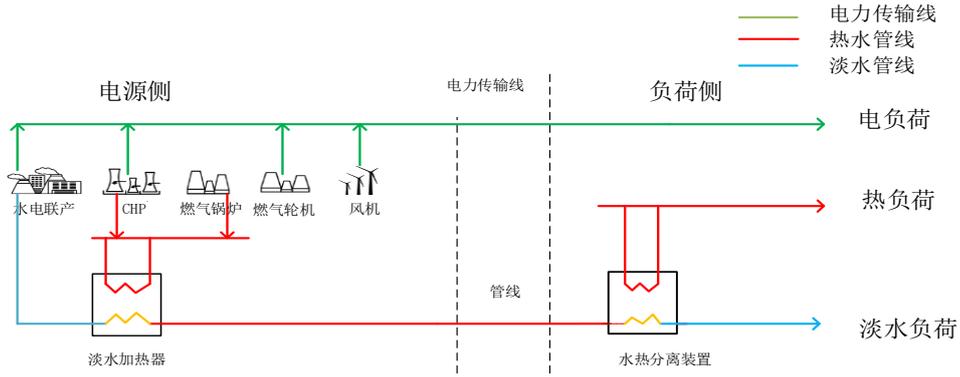

图 2 居民园区系统结构图

Fig. 2 System structure of residential community

本研究构建的多园区 IES 包括工业园区、商业园区及居民园区，并主要考虑用户的电负荷、热负荷以及淡水负荷等多种负荷需求。含多种功能园区的 IES 的结构见图 1。

由图 1 可知，不同园区之间存在着电能和热能两种能源形式的交互；而淡水的生产与消耗则在各自的园区内完成，不参与园区之间的互动。每个园区内部结构与组成要素基本相同，图 2 中以居民园区为例行了展示。

如图 2 所示，每个园区内都包括风力发电机组、热电联产（Combined Heat and Power，CHP）机组、水电联产机组、快速反应的燃气锅炉和燃气轮机等，充分考虑了对可再生能源的利用和对多种负荷需求的供给。其中，电能通过电力传输线传送到用户侧，而淡水和热则通过"水热同传"的模式，共用一个传输管道传送到用户侧。

## 1.2 水电联产机组

### 1.2.1 水电联产机组结构和原理

水电联产机组是火电机组与海水淡化装置相结合的一种水、电同产的高效联产设备[17]。反渗透海水淡化装置由于其能耗低、体积小，是目前主流的海水淡化设备。火电机组与海水淡化装置有很强的互补特性，水电联产机组结构和运行原理如图 3 所示[18]：

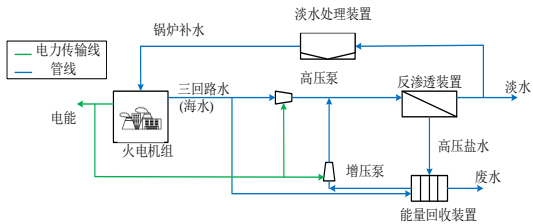

图 3 水电联产机组结构图

Fig. 3 Structure diagram of hydropower cogeneration unit

火电机组三回路冷却水经高压泵加压后推入反渗透装置，此后一部分海水通过渗透膜后变成淡水，剩余的高浓度海水被能量回收装置回收，与部分海水混合后，由增压泵再次送入反渗透淡化单元。其中，高压泵及增压泵所需电能由火电机组供应，部分淡水经过处理后可作为火电机组的锅炉补水。由火电机组和海水淡化装置联合运行的水电联产机组，通过对资源的互补利用，实现了淡水和电能的高效联产。

### 1.2.2 水电联产机组的能耗特性

渗透膜两侧的渗透压 $P_a$ 与海水浓度呈正比例关系[19]：

$$P_a = \lambda \vartheta_t = \frac{\lambda \vartheta_0}{1-B} \quad (1)$$

式中：$\lambda$ 为渗透压系数，$\vartheta_t$ 为当前反渗透装置内的海水盐度，$\vartheta_0$ 为海水初始盐度，$B$ 为回收率，即产出的淡水体积与入口海水体积的比值。

$$B = \frac{V_o}{V_i} \quad (2)$$

式中：$V_o$ 为反渗透装置所产淡水体积，$V_i$ 为反渗透装置所进入的海水体积。从而，水电转换系数 $N$，即生产单位体积淡水所需电量为：

$$N = \frac{1}{V}\int_0^V P_a dV = \lambda \vartheta_0 \frac{1}{B}\ln(\frac{1}{1-B}) \quad (3)$$

由上式可知，生产单位体积淡水所需电量 $N$ 与回收率 $B$ 之间有显著的凹函数关系。

因此，选择最适合的回收率，可以使 $N$ 值最小。$N$ 值最小状态下海水淡化系统所消耗的电功率为：

$$P_{ro} = Q \cdot w_{cwp} \quad (4)$$

式中：$P_{ro}$ 为海水淡化系统所需电功率，$w_{cwp}$ 为淡水生产速率。从而，水电联产机组对外输出的电功率为：

$$p_{cwp} = p_{tp} - p_{ro} \quad (5)$$

式中：$p_{cwp}$ 为水电联产机组对外输出的电功率，$p_{tp}$ 为火电机组的电功率。

### 1.3 水热同传系统

"水热同传"模型的物理结构如图 2 所示。在发电侧，CHP 机组和燃气锅炉产生的热能通过淡水加热装置将淡水加热，而后通过输水管道将承载热量的淡水传输至负荷侧实现物质能量共同传输，负荷侧的水热分离装置实现淡水与热能的分离然后分别供给热用户与淡水用户。

一般来说，热量输送系统的效率主要受两个方面的影响：加热器和水热分离装置的热效率，以及管线的热辐射效应引起的热衰减。淡水加热装置和水热分离装置与传统热网中的换热站类似，具有很高的传热效率，淡水加热装置的核心元件的基本结构原理如图 4 所示。

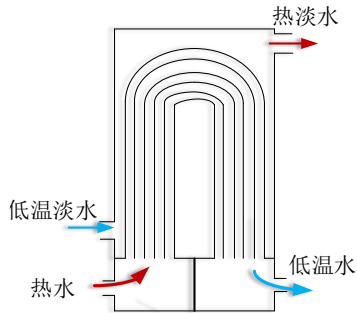

**图 4 淡水加热器结构原理示意图**

**Fig. 4 Structure diagram of fresh water heater**

如图 4 所示，来自 CHP 机组和燃气锅炉的热水在淡水加热器将热量传递给来自水电联产机组的低温淡水，使其成为携带热能的热淡水。由于加热器和水热分离装置热效率较高且相对固定，故主要对管线的热辐射效应带来的热衰减进行讨论。本文采用的是输热网的温度调节模式[20]，在时段 $t$ 内管道输送的热能 $H_t$ 为：

$$H_t = C^W G \left( T_t^{sw} - T_t^{rw} \right) \quad (6)$$

式中：$C^W$ 为水的比热容，$G$ 为时间 $t$ 内水热同传系统传输的水的质量，$T_t^{sw}$ 为流入水热分离装置的热淡水温度，$T_t^{rw}$ 为流出水热分离装置的低温淡水的温度。根据稳态传热基本定理，长度为 $L$ 的传输管线的热量损失 $\Delta H_t$ 可由如下公式计算得来[21]：

$$\Delta H_t = 2\pi \frac{T_t^{sw} - T_t^e}{\gamma} L \quad (7)$$

式中：$T_t^e$ 为管道外的环境的平均温度；$\gamma$ 为热介质与周围环境之间的平均热阻。因此，每个调度时段所需提供的热量为热负荷与传输损失之和，即：

$$H_t = h_{load} + \Delta H_t \quad (8)$$

由于水的比热容很大，完全可以满足作为供热传输载体的需求。"水热同传"这一物质能量共同传输模式，仅需要一条传输通道，极大的减少了建设阶段的人力物力成本，且在运行期间节约了大量的淡水资源，缓解了海岛本身淡水资源短缺的问题。位于中国山东沿海属于国家电力投资集团的"水热同传"示范工程[22-23]，经过一个供暖季的实际运行，验证了该模式的可行性与优越性，为淡水资源紧张地区的供暖、供水探索了一条行之有效的新路径。

关于 CHP 机组、燃气轮机以及燃气锅炉，在先前工作中已详细分析了其能耗特性并构建了出力模型，在此不再赘述[23]。

## 2 多园区综合能源系统调度模型

### 2.1 目标函数

多园区 IES 优化调度的目标是基于多种园区不同的负荷特性，利用其互补特性调整调度方案，在不同园区之间进行合理的能量交互，从而使系统的总运行费用最低。多园区 IES 优化调度的目标为：

$$F = \min C_E = C_1 + C_2 + C_3 \quad (9)$$

式中：下标 1，2，3 分别代表工业园区、商业园区和居民园区。本研究仅考虑一次能源的购买成本，不同的功能园区所包含的设备是一致的，因此其运行成本的数学表示也是一致的。本节以工业园区为例，对其运行成本的数学表示予以展开。

$$C_1 = \sum_1^{n_a} M_a \left( p_{1chp}, h_{1chp} \right) + \sum_1^{n_d} M_d \left( p_{1gt} \right) \\ + \sum_1^{n_b} M_b \left( p_{1cwp}, w_{1cwp} \right) + \sum_1^{n_e} M_e \left( h_{1gb} \right) \quad (10)$$

式中：$M_a$、$M_b$、$M_d$、$M_e$ 分别为 CHP 组、水电联产机组、燃气轮机、燃气锅炉的成本系数；$n_a$、$n_b$、$n_d$、$n_e$ 分别为各类设备的数量，$p_{1chp}$、$h_{1chp}$ 分别为 CHP 机组的电功率

和热功率，$p_{1cwp}$、$w_{1cwp}$ 分别为水电联产机组的电功率和淡水产水速率，$p_{1gt}$ 为燃气轮机的电功率，$h_{1gb}$ 为燃气锅炉的热功率。各个设备的运行成本展开的数学表达式为：

$$M_a\left(p_{1chp},h_{1chp}\right) = \rho_{gas}\left(\frac{p_{1chp}}{\eta_a \cdot HHV}\right)\Delta t + \rho_{gas}\left(\frac{h_{1chp}}{\eta_a \cdot HHV}\right)\Delta t \quad (11)$$

$$M_b\left(p_{1cwp},w_{1cwp}\right) = \rho_{gas}\left(\frac{p_{1cwp}}{\eta_b \cdot HHV}\right)\Delta t + \rho_{gas}\left(\frac{w_{1cwp}\cdot Q}{\eta_b \cdot HHV}\right)\Delta t \quad (12)$$

$$M_d\left(p_{1gt}\right) = \rho_{gas}\left(\frac{p_{1gt}}{\eta_d \cdot HHV}\right)\Delta t \quad (13)$$

$$M_e\left(p_{1gb}\right) = \rho_{gas}\left(\frac{h_{1gb}}{\eta_e \cdot HHV}\right)\Delta t \quad (14)$$

式中：$\rho_{gas}$ 为单位热值天然气价格，$\eta$ 为各设备的效率。文中商业和居民园区的运行成本计算方法与工业园区类似，故不予赘述。

### 2.2 约束条件

对于多园区 IES，主要考虑功率平衡约束、设备运行限值约束以及各个园区之间功率交互的约束。

1）功率平衡约束

对于多园区 IES，优化运行的前提是要保证各个园区的多种负荷需求得到满足，因此多园区 IES 的功率平衡约束条件为：

$$\begin{cases} p_{ichp} + p_{icwp} + p_{igt} + p_{iwt} + p_{ijh} = p_{iload} \\ h_{ichp} + h_{igb} + h_{ijh} = h_{iload} \quad i \in 1,2,3 \\ w_{icwp} = w_{iload} \end{cases} \quad (15)$$

式中：$p_{ijh}$ 为 $i$ 园区与其他园区交互的电功率，当从其他园区获得电功率时，$p_{ijh}$ 为正，当向其他园区传输功率时，$p_{ijh}$ 为负；$h_{ijh}$ 为 $i$ 园区与其他园区交互的热功率，当从其他园区获得热功率时，$h_{ijh}$ 为正，当向其他园区传输热功率时，$h_{ijh}$ 为负。值得注意的是，各个园区的负荷以及 RE 出力情况是不确定的。此外，对于不同园区之间的功率交互应当守恒，即交换功率的代数和为 0。

$$p_{1jh} + p_{2jh} + p_{3jh} = 0 \quad (16)$$
$$h_{1jh} + h_{2jh} + h_{3jh} = 0 \quad (17)$$

2）设备运行限值约束

对于 CHP 机组，其输出电功率和热功率之间的耦合关系称为"热电特性"。热功率与电功率的比值称为热电比 $b$，根据热电比 $b$ 是否可以变化 CHP 机组可以分为定热电比机组和变热电比机组。本研究选取的 CHP 机组为常见的背压式机组，这是一种变热电比机组，通过将热电联产机组的电功率 $p_{chp}$ 和热电比 $b$ 控制在其运行限值内，从而保证热电联产机组不会超过其运行极限。热电联产机组的运行限制约束可由如下公式（18）和（19）表示[24]：

$$p_{chp\min} \leq p_{chp} \leq p_{chp\max} \quad (18)$$
$$b_{\min} \leq b \leq b_{\max} \quad (19)$$

式中：$p_{chp\max}$ 和 $p_{chp\min}$ 分别为热电联产机组电功率的上限和下限，$b_{\max}$ 和 $b_{\min}$ 分别是热电比的上限和下限。

对于水电联产机组，其实质为利用电能驱动泵体对海水加压使其穿过渗透膜实现海水淡化，因此进行海水淡化所需的电功率与向外输出的电功率之和不应超过水电联产机组的功率极限，故而：

$$p_{tp\min} \leq p_{cwp} + Qw_{cwp} \leq p_{tp\max} \quad (20)$$

式中：$p_{tp\max}$ 和 $p_{tp\min}$ 分别为构成水电联产机组的火电机组部分电功率的上下限。

对于燃气轮机和燃气锅炉应当运行在其允许的最大最小功率之间，故而其运行限制分别需要满足以下约束[25-26]：

$$p_{gt\min} \leq p_{gt} \leq p_{gt\max} \quad (21)$$
$$h_{gb\min} \leq h_{gb} \leq p_{gb\max} \quad (22)$$

式中：$p_{gt\max}$ 和 $p_{gt\min}$ 分别为燃气轮机电功率的上下限；$h_{gb\max}$ 和 $h_{gb\min}$ 分别为燃气锅炉热功率的上下限。此外，在多园区 IES 的调度模型中，还应考虑不同园区之间的功率交互约束[27]：

$$|p_{1jh}|+|p_{2jh}|+|p_{3jh}| \leq 2p_{jh\max} \quad (23)$$
$$|h_{1jh}|+|h_{2jh}|+|h_{3jh}| \leq 2h_{jh\max} \quad (24)$$

式中：$p_{jh\max}$ 和 $h_{jh\max}$ 分别为某一时间段内，不同园区之间电功率和热功率交互的最大限值。

## 3 基于深度强化学习的多园区综合能源系统优化调度

### 3.1 多智能体深度强化学习算法

DRL 的核心思想是让智能体通过试错并利用反馈以不断地改进行动决策[28]。在

DRL 算法中，智能体以获取最大奖励为目标，通过不断地与环境交互来探索最优策略。智能体与环境的互动是一种 MDP[29]，传统的 MDP 由四个元素组成，即（$S$, $A$, $R$, $\pi$）组成[30]，其中 $S$ 为所有环境状态的集合；$A$ 为智能体所有可执行动作的集合；$R$ 为奖励函数；$\pi$ 为智能体的策略集。

本研究采用的多智能体近端策略优化（Multi-agent Proximal Policy Optimization，MAPPO）算法是在近端策略优化算法（Proximal Policy Optimization，PPO）算法的基础上发展出来的一种多智能体变种，它是一种基于 actor-critic 框架的多智能体深度强化学习（Multi-agent Deep Reinforcement Learning, MADRL）算法[31]，适用于解决多个主体之间的交互决策问题。PPO 算法包括三个网络，两个 actor 网络和一个 critic 网络。两个 actor 网络分别拟合的是智能体的新旧行动策略，即 $\pi(a_t|s_t)$ 和 $\pi_{old}(a_t|s_t)$。新旧策略之间的变化比例为 $r$，它用来反映新、老策略之间的差异[32]。

$$r = \frac{\pi(a_t|s_t)}{\pi_{old}(a_t|s_t)} \quad (25)$$

通过将变化比例 $r$ 限制在一定的范围内，可以降低 PPO 算法对学习率的敏感程度，从而使得策略更加稳定。不同于传统的 PPO 算法，MAPPO 算法中有多个相对独立的智能体，其网络结构框架如图 5 所示[33]。

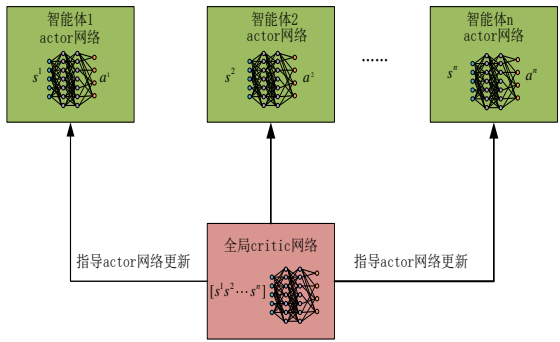

**图 5 MAPPO 网络结构框架**

**Fig. 5 MAPPO network structure framework**

由图 5 可知，MAPPO 中每个智能体有一个独立的 actor 网络，各智能体共享一个共同的 critic 网络。这意味着所有智能体可感知和评估全局状态的中心值函数，并且能够协同地朝着一个共同的目标前进、进行探索。本文所用的 MAPPO 算法是一种集中式训练，分布式执行的混合结构的多智能体 DRL 算法，对于此类各个园区之间属于合作关系的多园区 IES 调度问题非常适用。

### 3.2 基于深度强化学习的调度框架

DRL 的基本结构框架包括环境状态集合 $S$、智能体可以选择的动作集合 $A$ 和指导智能体进行探索的奖励函数 $R$。结合所构建的多园区 IES，本节将 2.2 节中的调度模型转化为多智能体 DRL 框架。在某一时段内，智能体感知到的环境状态变量主要为各个园区的负荷情况和可再生能源的出力情况，因此智能体探索的全局环境状态空间为：

$$S = [p_{1load}, h_{1load}, w_{1load}, p_{1wt},$$
$$p_{2load}, h_{2load}, w_{2load}, p_{2wt}, \quad (26)$$
$$p_{3load}, h_{3load}, w_{3load}, p_{3wt},]$$

智能体可以选择的动作为各个园区中设备的出力情况以及各个园区之间的功率交换，因此智能体的动作集合为：

$$A = [p_{1chp}, h_{1chp}, p_{1cwp}, w_{1cwp}, p_{1gt}, h_{1gb}, p_{1jh}$$
$$p_{2chp}, h_{2chp}, p_{2cwp}, w_{2cwp}, p_{2gt}, h_{2gb}, p_{2jh} \quad (27)$$
$$p_{3chp}, h_{3chp}, p_{3cwp}, w_{3cwp}, p_{3gt}, h_{3gb}, p_{3jh}]$$

文中多园区 IES 优化调度的目标是在满足功率平衡约束和各类运行限值的约束的基础上，实现各园区之间的协调互动，以寻求最低运行成本。故而，奖励函数设置为：

$$R = -\frac{C_E + \sum_{i=1}^{n} D_i}{Z} + U \quad (28)$$

式中：$U$ 为一个合适的正数，用来避免奖励函数一直为负数；$M$ 为不满足各类设备约束限值的惩罚项；$D_i$ 为各个园区电源、负荷之间的不平衡功率，主要包括电能需求与供给之间的不平衡，热能需求与供给之间的不平衡，淡水供给与需求之间的不平衡。$D_i$ 可通过下式来具体描述：

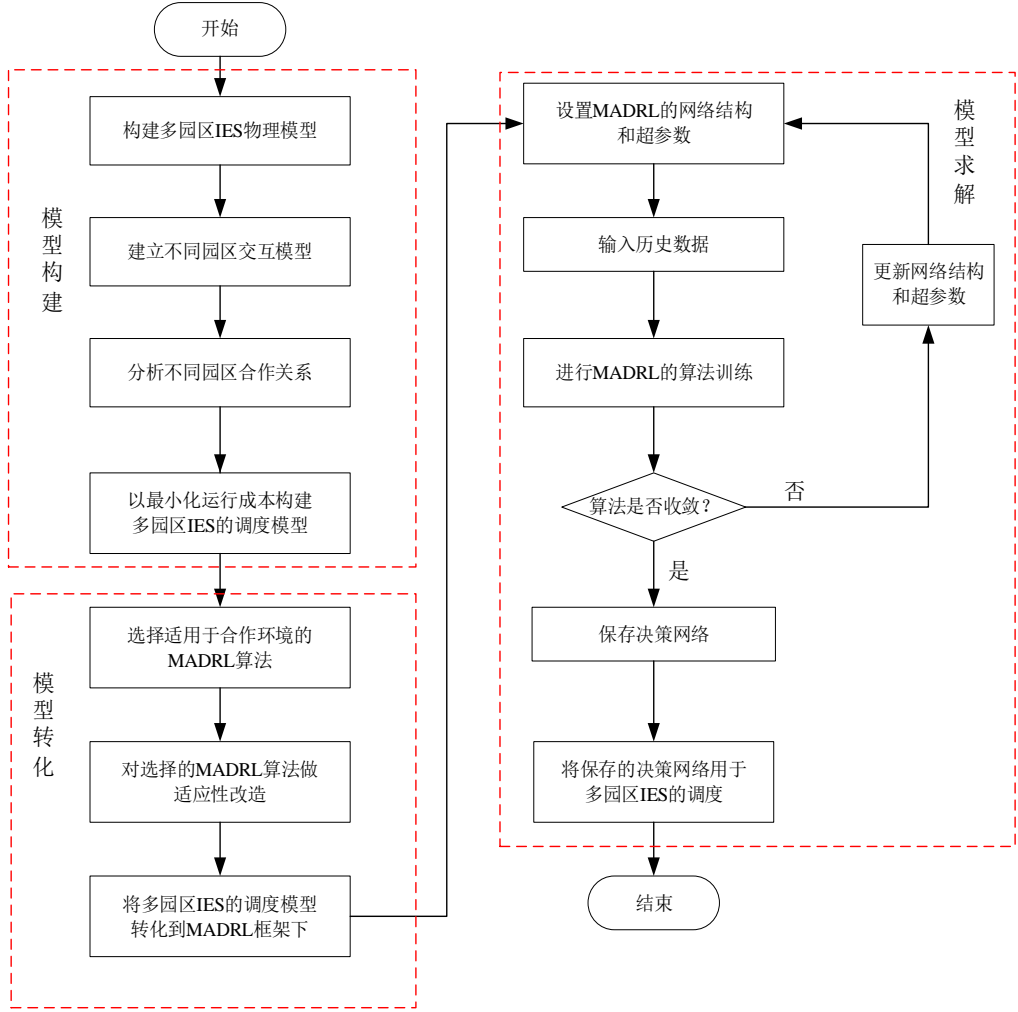

**图 6 所提方法流程图**

**Fig. 6 Flow chart of the proposed method**

$$D_i = |p_{ichp} + p_{icwp} + p_{igt} + p_{iwt} + p_{ijh} - p_{iload}|$$
$$+ |h_{ichp} + h_{igb} + h_{ijh} - h_{iload}| + \quad i \in 1,2,3 \quad (29)$$
$$Q|w_{icwp} - w_{iload}|$$

## 4 求解流程

文中所提的关于含多重不确定性的多园区海岛 IES 的优化调度流程如图 6 所示，主要步骤总结如下：

步骤 1：构建多园区 IES 物理模型；

步骤 2：建立多园区电、热能源交互模型；

步骤 3：以最小化运行成本构建多园区海岛 IES 调度模型；

步骤 4：选择适用的 MADRL 算法并进行适应性改造；

步骤 5：研究 MAPPO 算法的框架和网络结构；

步骤 6：将多园区 IES 的调度模型转化到 MAPPO 算法框架下；

步骤 7：设置 MAPPO 算法的网络结构和超参数；

步骤 8：对 MAPPO 算法进行训练；

步骤 9：如果算法收敛，则保存该决策网络；否则，转向步骤 8，调整算法结构和超参数。

步骤 10：将保存的决策网络用于多园区 IES 的调度；

步骤 11：对调度结果进行分析。

## 5 算例分析

### 5.1 测试系统介绍

为了验证本文所提的基于 MAPPO 算法的多园区 IES 的调度模型的有效性，本节构建了一个多园区的 IES 进行测试，系统结构和原理见图 7。

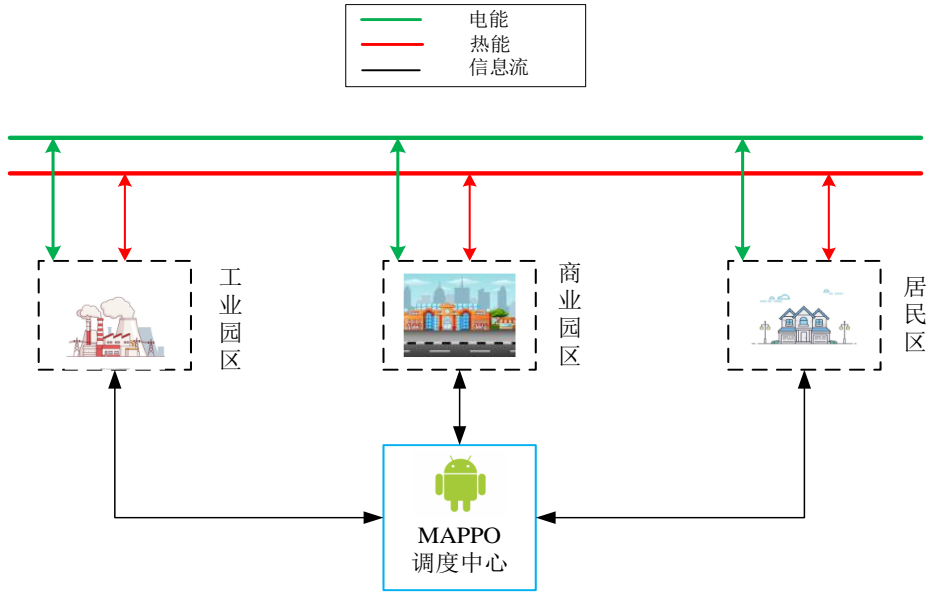

**图 7 测试系统结构图**

**Fig. 7 Test system structure diagram**

表 1 各主要设备的运行参数

Table 1 Operation parameters of each main device

| 设备 | 参数 | 符号 | 数值 |
| --- | --- | --- | --- |
| 电热联产机组 | 电功率上限 | $p_{chp\,max}$ | 5000 kW |
|  | 电功率下限 | $p_{chp\,min}$ | 1000 kW |
|  | 热电比 | $b$ | [0, 1.4] |
| 水电联产机组 | 功率上限 | $p_{tp\,max}$ | 5000 kW |
|  | 水电转换系数 | $N$ | 8 |
|  | 最大产水体积 | $V$ | 500 m$^3$ |
|  | 压力参数 | $\lambda$ | 4.17(MPa·L)/mol |
|  | 初始海水浓度 | $\vartheta_0$ | 0.6 mol/L |
| 燃气轮机 | 电功率上限 | $p_{gt\,max}$ | 3000 kW |
| 燃气锅炉 | 热功率上限 | $h_{gb\,max}$ | 3000 kW |

如图 7 所示，不同园区在 MAPPO 算法的指导下，进行热能和电能的交互。每个园区中电源设备包含 CHP 机组、水电联产机组、燃气轮机和燃气锅炉，以及分布式风力发电机组,这些设备的参数如表1参数所示。另外，不同园区之间可以进行电能和热能交互的最大功率限值 $p_{jh\,max}$ 和 $h_{jh\,max}$ 均为 1000kW。本测试系统的调度周期为 24 小时，每个调度时段为 1 小时。参考相关文献[34-35]，文中 MAPPO 算法的结构和超参数设置如下：本算法的所有智能体各有两条 actor 网络，所有的智能体共享一条 critic 网络，每个网络都包含两个隐藏层，每层包含 128 个神经元，actor 网络的学习率设置为 0.00004，而 critic 网络的学习率设置为 0.0004，批量调取进行训练的数据大小为 96 个，训练回合数为 25000 次。文中采用冬季北方某多园区 IES 中不同园区的负荷数据作为原始数据，将该原始数据进行容量适应性处理后作为 MAPPO 算法的训练数据，训练结束后，抽取某一典型日的负荷数据进行测试分析。MAPPO 算法的训练过程见图 8。

由图 8 可知，算法经过约 20000 个回合训练后平均奖励趋于平稳，这表明算法已收敛，智能体获得了最优的 IES 调度策略。

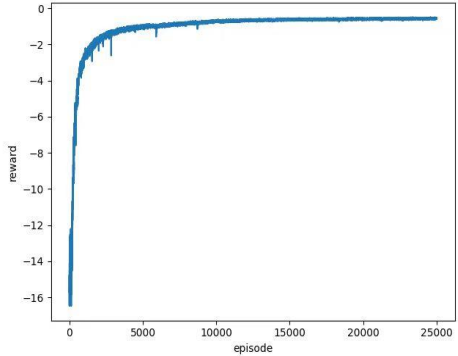

**图 8 训练过程奖励变化曲线**

**Fig. 8 Reward curve during training**

## 5.2 调度结果分析

为了验证文中所提的基于MAPPO的调度方案是否可以自主地学习到不同园区的负荷特征，并利用这些特征中的互补特性指导园区之间合理的能量交互，本节构建了两个典型运行模式进行对比分析。

模式一：基于单智能体 DRL 的调度方案，各个园区独立运行，不进行能量的交互。

模式二：基于多智能体 DRL 的调度方案，各个园区之间进行电能、热能交互。

1）运行成本分析

为了验证所提调度方案的优越性，将两种模式的运行成本做了整理，并汇总成表2。

**表 2 不同模式运行成本对比表**

**Table 2 Comparison table of operating costs in different modes**

|  | 模式一 | 模式二 |
|---|---|---|
| 工业园区成本/¥ | 132376.7 | **128167.5** |
| 商业园区成本/¥ | 95474.5 | **90656.7** |
| 居民园区成本/¥ | **90983.2** | 94564.6 |
| 总成本/¥ | 318834.4 | **313388.8** |

由表 2 可知，相对于模式一中各个园区独立运行，模式二联合运行后各个园区的运行成本均发生了变化。在模式二中，工业园区的运行成本为 128167.5 元，比模式一的运行成本低了 4209.2 元；商业园区的运行成本为 90656.7 元，比模式一的运行成本低了 4817.8 元；居民园区的运行成本为 94564.6 元，比模式一的运行成本高了 3581.4 元。这说明多园区联合调度运行后，不同园区之间进行了能量流动。对于 IES 的总运行成本，模式二为 313388.8 元，比模式一低了 5445.6 元，这表明多园区系统联合运行后，MADRL 算法根据不同园区之间的负荷特性合理安排各个园区内设备的出力，实现了不同园区之间的合理能量流动，从而优化了总体调度方案并降低了运行成本。

2）调度动作分析

应用基于MAPPO的多园区IES联合运行调度模型的模式二的运行成本相较于模式一明显降低。为了分析具体原因，本节将模式一与模式二在不同时段关于电能和热能的调度动作进行了汇总整理，结果如图9和图10所示。

图 9（a）、（b）、（c）分别为工业园区、商业园区和居民园区在模式一下各个园区独立运行时的电力调度子图。图 10（d）、（e）、（f）分别为工业园区、商业园区和居民园区在模式二下各个园区联合运行时的电力调度子图。

由图 9 可知，无论是各个园区独立运行的模式一，还是多园区联合运行的模式二，基于 DRL 的调度方案都可以很好的满足各个园区的负荷需求。总体来看，多园区联合运行后，商业园区的电功率向工业园区和居民园区流动。对比图 9 的（b）和（e），商业园区在独立运行时夜间消纳的可再生能源功率远低于联合运行后，这是因为商业园区在夜间的电负荷较小，加之 CHP 机组由于热电耦合具有一定的强迫出力。多园区联合运行后，MAPPO 算法安排将商业园区无法消纳的风机功率转移到工业园区，同时降低工业园区水电联产机组的电功率，以实现对这些可再生能源的消纳。商业园区和工业园区的这种电能转移，是 MAPPO 算法在学习到不同园区的互补负荷特性基础上进行合理调度决策，不仅可以实现对 RE 的完全消纳，又降低了多园区 IES 的整体运行费用。

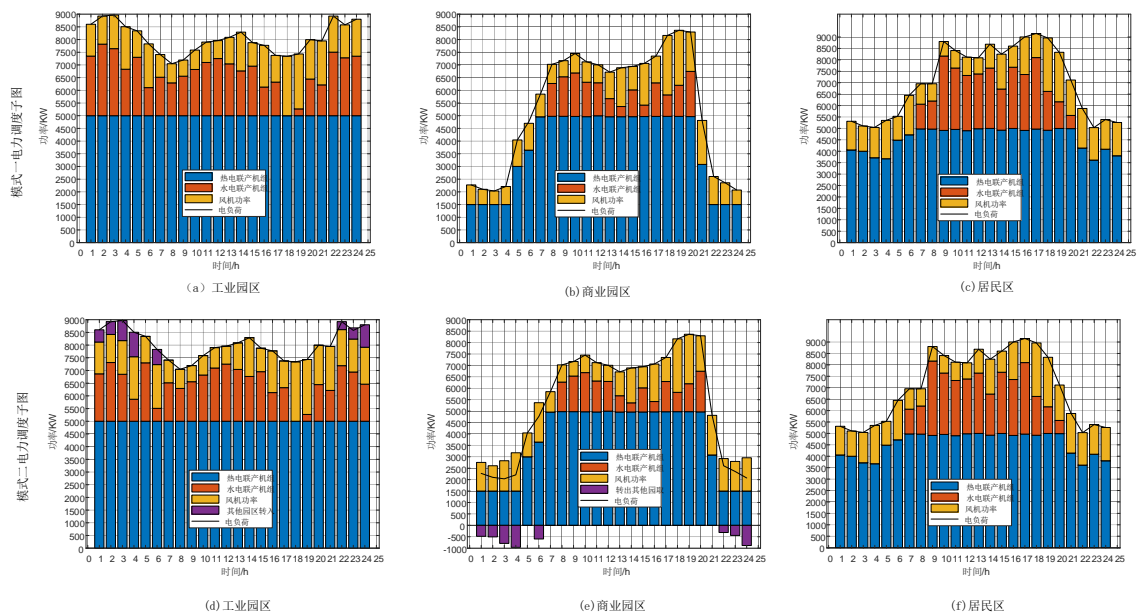

图 9 不同模式各个园区电力调度子图

Fig. 9 Power dispatching subgraph of each community in different modes

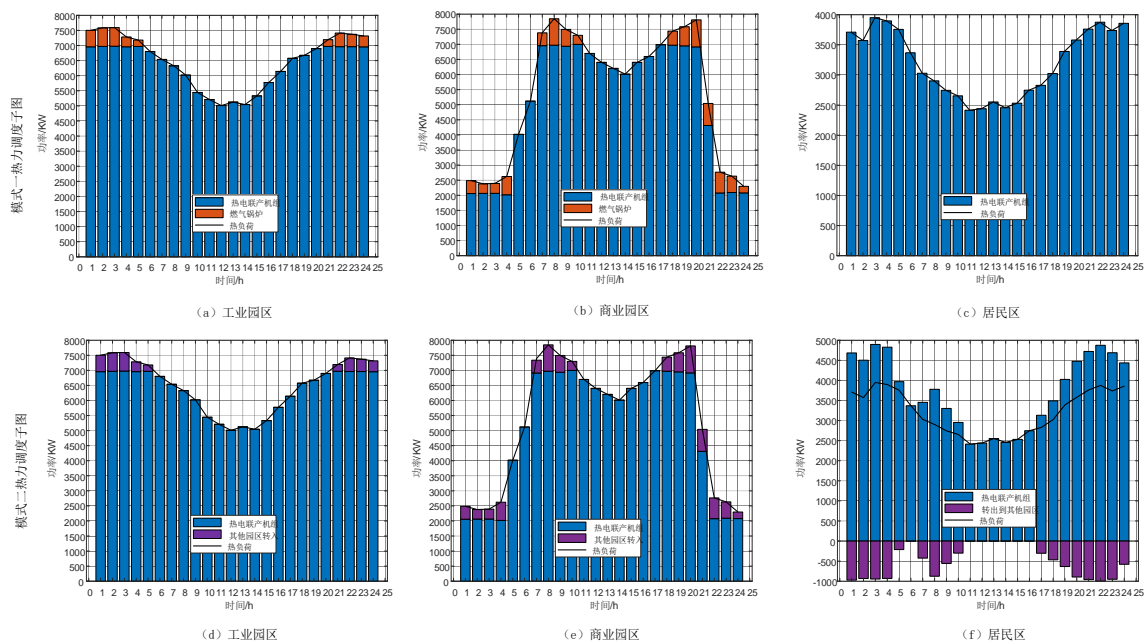

图 10 不同模式各个园区热力调度子图

Fig. 10 Heat dispatching subgraph of each community in different modes

图 10 中（a）、（b）、（c）分别为工业园区、商业园区和居民园区在模式一中各自独立运行时的热力调度子图；图 9 中（d）、（e）、（f）分别为工业园区、商业园区和居民园区在模式二中多园区联合运行时的热力调度子图。总体来看，各园区独立运行时，工业园区和商业园区在负荷较高的时段需要 CHP 机组和燃气锅炉共同运行来满足热负荷需求。这是因为在这些热负荷需求较高的时段，CHP 机组已经达到热功率上限，无法独立承担园区内全部的热负荷。然而，在多园区联合运行时，MAPPO 决策网络安排居民园区的 CHP 机组在夜间时段提升热功率并向相应的园区输送热能，以承担更多的热负荷。这是因为 CHP 机组单位热功率成本较低于燃气锅炉，通过用居民园区的单

位运行成本较低的 CHP 机组替代其他园区单位运行成本较高的燃气锅炉,虽然会增加居民园区的运行成本,但整体上会降低多园区 IES 的总运行成本。由图 10(d)、(e)、(f)可知,各个园区的燃气锅炉全部退出运行,所有的热负荷均由运行成本较低且效率较高的 CHP 机组承担。特别在图 10(b)中可以观察到,尽管商业园区在夜间时段的热负荷很低,仍需要启动燃气锅炉以满足热负荷需求。这是由于商业园区在夜间时段的电负荷很低,由于 CHP 机组的热电耦合关系,其热功率极限也很低,难以单独满足这种负荷需求。

通过对电力、热力调度动作的分析,可以发现在多园区 IES 联合运行后,基于 MAPPO 的调度决策网络可以很好地学习到不同园区之间的负荷互补特性。在此基础上,该算法能够做出合理的调度决策,安排各园区之间进行合理的能量流动,从而实现多园区 IES 总体运行成本降低。

3)可再生能源消纳分析

为了具体分析基于 MAPPO 的多园区 IES 对 RE 出力消纳的提升情况,本节以商业园区为例,将其在不同运行模式下 RE 出力的消纳情况进行了分析。不同模式下 RE 消纳和 RE 出力情况如图 11 所示。

由图 11 可知,模式一在夜间时段存在对 RE 大量的浪费,而模式二几乎在任何时段的 RE 消纳情况都要优于模式一,并且模式二中消纳的 RE 与 RE 实际出力的情况相一致。这说明基于 MAPPO 的多园区联合调度模型可以提升对 RE 的消纳,实现对其充分利用。

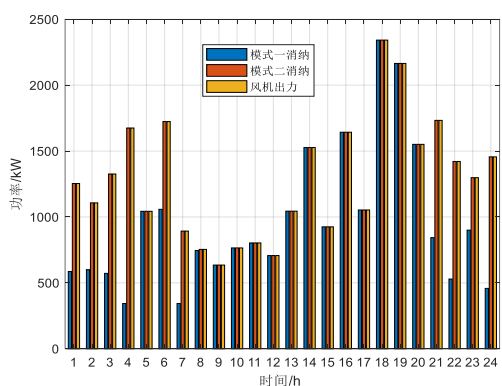

**图 11 不同模式下可再生能源消纳情况**
**Fig. 11 Consumption of renewable energy under different modes**

## 6 结论

本文提出了一种利用多智能体 DRL 算法实现多园区 IES 协调运行的优化调度模型。该模型通过学习不同园区的负荷特征,并协调不确定环境下各个园区之间的能量交互,实现系统整体的优化运行。在多园区模式下,综合能源系统的各园区以及各能源子系统的耦合关系更加复杂,对调度系统的协调处理能力要求成倍提高。仿真结果表明,所提方法可以很好地捕捉不同园区的负荷特性,并利用其互补特性来协调不同园区之间的能量交互,避免了对复杂耦合关系进行建模,具有良好的经济效益和环保效益。

## 7 参考文献

**作者简介：**

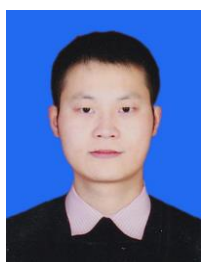

李扬(1980)，男，教授，博士生导师，通信作者，主要研究方向为：综合能源系统优化调度、电力系统稳定评估，E-mail： liyang@neepu.edu.cn；

马文捷(1999)，男，硕士研究生，主要研究方向为：人工智能在综合能源系统中的应用；

卜凡金(1996)，男，硕士，主要研究方向为：人工智能在综合能源系统中的应用；

杨震(1992)，男，硕士，工程师，主要研究方向为：电力系统优化运行与控制；

王彬(1995)，男，硕士，工程师，主要研究方向为：电力系统优化运行与控制；

韩猛(1996)，男，硕士，工程师，主要研究方向为：电力系统优化运行与控制。

(编辑　***)